\documentclass[12pt]{amsart}
\usepackage{amsmath,amssymb,cite,enumerate,graphicx,hyperref}

\hypersetup{pdftex,colorlinks=true,allcolors=blue}
\usepackage[all]{hypcap}

\usepackage[foot]{amsaddr}
\usepackage[sc]{mathpazo}

\numberwithin{equation}{section}

\DeclareMathAlphabet\mathsfbi{T1}{phv}{b}{it}

\newcommand\dif{\,\mathrm{d}}
\newcommand\EE{\mathbb E}
\newcommand\BV{\boldsymbol}
\newcommand\BM{\mathsfbi}
\newcommand\parderiv[2]{\frac{\partial #1}{\partial #2}}
\renewcommand\div{\mathrm{div}}
\newcommand\trace{\mathrm{tr}}

\newcommand\ack{\subsection*{Acknowledgment}}

\begin{document}

\author[Rafail V. Abramov]{Rafail V. Abramov}

\address{Department of Mathematics, Statistics and Computer Science,
University of Illinois at Chicago, 851 S. Morgan st., Chicago, IL 60607}

\email{abramov@uic.edu}

\title[Gas near a wall: Reduced viscosity and the Knudsen layer]{Gas
  near a wall: a shortened mean free path, reduced viscosity, and the
  manifestation of a turbulent Knudsen layer in the Navier-Stokes
  solution of a shear flow}

\begin{abstract}
For the gas near a solid planar wall, we propose a scaling formula for
the mean free path of a molecule as a function of the distance from
the wall, under the assumption of a uniform distribution of the
incident directions of the molecular free flight. We subsequently
impose the same scaling onto the viscosity of the gas near the wall,
and compute the Navier-Stokes solution of the velocity of a shear flow
parallel to the wall. This solution exhibits the Knudsen velocity
boundary layer in agreement with the corresponding Direct Simulation
Monte Carlo computations for argon and nitrogen. We also find that the
proposed mean free path and viscosity scaling sets the second
derivative of the velocity to infinity at the wall boundary of the
flow domain, which suggests that the gas flow is formally turbulent
within the Knudsen boundary layer near the wall.
\end{abstract}

\maketitle

\section{Introduction}

An accurate description of the Knudsen velocity boundary layer is a
long-standing problem in the study of
microflows~\cite{Sone,ZhaMenWei}. Guo et al.~\cite{GuoShiZhe} proposed
to scale the viscosity of the gas near a wall proportionally to the
corresponding mean free path scaling (as the viscosity is generally
proportional to the mean free path of a gas molecule \cite{ChaCow}),
where the mean free path scaling near the wall was proposed by
Stops~\cite{Stops}. In~\cite{Stops}, the scaling implied that the
distribution of the free flight directions was proportional to the
cosine of the angle between the flight direction and the direction of
the shortest distance to the wall (called the ``cosine law''
in~\cite{Stops}). It appears, however, that a uniform distribution of
the molecular flight directions was not addressed so far in the given
setting.

In the current work we propose a new scaling for the mean free path of
a gas molecule near a solid planar wall, which is based on a uniform
distribution of the incident molecular flight directions. We start
with the probability distribution of the length of the free flight of
a molecule constrained by a general solid obstacle, and, for the
special case of a solid planar wall, we derive an exact scaling
formula for the mean free path as a function of the distance from the
wall. The same scaling subsequently applies to the viscosity of the
gas, since the viscosity is proportional to the mean free
path~\cite{ChaCow}. We then compute the Navier-Stokes solution of the
velocity of the shear flow parallel to the wall, and observe the
emergence of the Knudsen boundary layer. We find this boundary layer
to be quite similar to the one produced by the Direct Simulation Monte
Carlo (DSMC) computations for argon and nitrogen under normal
conditions. In comparison, the scaling based on the cosine law
from~\cite{GuoShiZhe,Stops} does not appear to model the Knudsen
boundary layer with comparable accuracy. Also, a surprising feature of
the viscosity scaling we propose is that it requires the second
derivative of the gas flow velocity to be infinite at the wall
boundary of the flow domain, which, from the perspective of the fluid
dynamics, suggests that the Knudsen boundary layer is formally
turbulent near the wall boundary.

\section{The behavior of the mean free path near a wall}

First, we consider the situation where a gas molecule flies freely
until it collides with another gas molecule (no obstacles are
present). We denote the conventional molecular mean free path in the
absence of obstacles as $\lambda_0$. Assuming that a collision is
random and can happen anytime with equal probability, here we model it
via a Poisson process, such that the length of the free flight $r$
until a collision is also a random variable. From the elementary
theory of random processes~\cite{GikSko} it follows that the
probability distribution of time until the next collision decays
exponentially with increasing time for a Poisson process model.
Observing that the speed of the free flight between collisions is
constant, we conclude that the probability distribution function
$f(r)$ of the length $r$ of the free flight until a collision also
decays exponentially with increasing $r$:
\begin{equation}
\label{eq:f}
f(r)=\frac 1{\lambda_0} e^{-r/\lambda_0}.
\end{equation}
The factor in front of the exponent above ensures that $f(r)$ is
normalized; also, the expectation $\EE$ of the length $r$ of the free
flight until a collision is precisely $\lambda_0$. The same
statistical distribution of the length of the free flight was used
in~\cite{Stops} (see the formula at the bottom of p. 686
in~\cite{Stops}), and, for the thermodynamic conditions close to
normal, is also corroborated by the direct molecular dynamics
simulations in~\cite{DonZhaRee} (see Fig.~3(a) in~\cite{DonZhaRee}).

Second, let an obstacle be placed along the direction of the flight at
the distance $r_0$ from the current position of the gas molecule, such
that the collision is guaranteed to occur before or at $r_0$. In this
case, the expectation $\EE$ of the length $r$ of the free flight until
a collision (either with another molecule, or, if none happens, with
the obstacle upon covering the distance $r_0$) becomes a function of
$r_0$ and is given by
\begin{equation}
\EE(r_0)=\int_0^{r_0} rf(r)\dif r+r_0\int_{r_0}^\infty f(r)\dif r
=\lambda_0\left(1-e^{-{r_0}/\lambda_0}\right).
\end{equation}
Above, the expectation integral is separated into two parts: the first
part computes the portion of the expectation of the mean free path for
the event where the molecule does not reach the obstacle (by colliding
with another molecule); the second part adds the portion corresponding
to the complementary event, given by the product of the distance to
the obstacle (since that would be the length of the free path for such
an event) and the probability of this event.

Third, let the direction of the flight be given by $\theta$ and
$\phi$, which are the azimuthal and polar angles, respectively, of a
suitable spherical coordinate system. Let us assume that the possible
directions of the molecular flight are distributed with the
probability density $g(\theta,\phi)$, which satisfies the
normalization condition
\begin{equation}
\int_0^{2\pi}\int_0^\pi g(\theta,\phi)\sin\phi\dif\phi\dif\theta=1.
\end{equation}
Also, let the corresponding distance to the obstacle depend on the
direction of the free flight, that is, $r_0=r_0(\theta,\phi)$. Then,
in order to compute the mean free path $\lambda$, we integrate
$\EE(r_0)$ against $g(\theta,\phi)$ as follows:
\begin{multline}
\label{eq:lambda1}
\lambda=\int_0^{2\pi}\int_0^\pi\EE(r_0(\theta,\phi))g(\theta,\phi)
\sin\phi\dif\phi\dif\theta=\\=\lambda_0\left(1-\int_0^{2\pi}
\int_0^\pi e^{-r_0(\theta,\phi)/\lambda_0}g(\theta,\phi)\sin\phi\dif
\phi\dif\theta\right).
\end{multline}

Fourth, let a planar wall be placed at the distance $d$ from the
origin along the ``north pole'' direction of the spherical coordinate
system, and let the ``southern'' hemisphere be free of
obstacles. Then, $r_0(\theta,\phi)$ is given by
\begin{equation}
\label{eq:x}
r_0(\theta,\phi)=\left\{\begin{array}{l@{\qquad}l} d/\cos\phi, &
0\leq\phi<\pi/2, \\ \infty, & \pi/2\leq\phi\leq\pi.
\end{array}\right.
\end{equation}
In this case, the formula in~\eqref{eq:lambda1} becomes
\begin{equation}
\label{eq:lambda2}
\lambda=\lambda_0\left(1-\int_0^{2\pi}\int_0^{\pi/2} e^{-d/(\lambda_0
  \cos\phi)}g(\theta,\phi)\sin\phi\dif\phi\dif\theta\right).
\end{equation}
Observe that above in~\eqref{eq:lambda2} the form of the directional
probability distribution $g(\theta,\phi)$ is unimportant for
$\phi>\pi/2$ (that is, for the flight directions escaping the
wall). Thus, in what follows, we only need to define $g(\theta,\phi)$
for the incident flight directions, that is, for
$0\leq\phi\leq\pi/2$. Next, we are going to consider two different
options for $g(\theta,\phi)$: the uniform distribution and the cosine
distribution.

\subsection{Uniform distribution of incident flight directions}

Here we assume that
\begin{equation}
  g_u(\theta,\phi)=\frac 1{4\pi}\quad\mathrm{for}\quad 0\leq\phi\leq
  \pi/2,
\end{equation}
that is, the distribution of the incident flight directions is
uniform. This seems to be a reasonable choice in the absence of a more
detailed statistical information on the flight directions in the
vicinity of a wall. Observe that the integral of $g_u$ over the
``northern'' hemisphere yields $1/2$, as, on average, the number of
molecules approaching the wall equals the number of molecules escaping
the wall.

The assumption of the invariance of $g$ over the directions (and, in
particular, the azimuthal angle) implies that the bulk of the gas is
moving uniformly in the chosen coordinate system (in particular, there
is no velocity shear). While not strictly true in the numerical
experiments which are to follow, we show below via a simple estimate
that the effect of the imposed velocity shear on the mean free path
scale is minuscule compared to the bulk average molecular speed for
normal conditions, and thus cannot substantially invalidate the
assumption above.

For the uniform distribution of the flight directions, the integral
in~\eqref{eq:lambda2} is evaluated as
\begin{equation}
\int_0^{2\pi}\int_0^{\pi/2} e^{-d/(\lambda_0 \cos\phi)}\frac 1{4\pi}
\sin\phi\dif\phi\dif\theta=\frac 12\left(e^{-d/\lambda_0}-\frac
d{\lambda_0} E_1(d/\lambda_0) \right),
\end{equation}
where $E_1$ is a standard notation for the exponential integral
\begin{equation}
\label{eq:expint}
E_1(x)=\int_x^{\infty}\frac{e^{-y}}y\dif y.
\end{equation}
Thus, we find that the mean free path $\lambda$ at the distance $d$
away from the wall is given by the following scaling formula:
\begin{equation}
\label{eq:lambda_uniform}
\lambda_u(d)=\lambda_0\left(1-\frac 12\beta_u(d/\lambda_0)\right),
\qquad\beta_u(x)=e^{-x}-x E_1(x).
\end{equation}

\subsection{Cosine distribution of incident flight directions}

What is to follow leads to the single-wall variant of the scaling
formula suggested by Stops~\cite{Stops} and tested by Guo et
al.~\cite{GuoShiZhe}, and thus we present it here for completeness.
Here we set
\begin{equation}
g_c(\theta,\phi)=\frac 1{2\pi}\cos\phi\quad\mathrm{for}\quad 0\leq\phi
\leq\pi/2,
\end{equation}
that is, the probability of the incident flight direction is
proportional to the cosine of the angle between the flight direction
and the direction of the shortest distance to the wall. Observe that
the integral of $g_c$ over the ``northern'' hemisphere equals $1/2$,
for the same reasons as for $g_u$ above. As for the uniform
distribution above, the independence on the azimuthal angle implies
that the effect of the velocity shear on the mean free path scale, if
present, is small in comparison to the bulk average speed of molecules
(which we show to be true for the numerical experiments below).

In this case, the integral in~\eqref{eq:lambda2} is evaluated as
\begin{multline}
\int_0^{2\pi}\int_0^{\pi/2} e^{-d/(\lambda_0 \cos\phi)}\frac 1{2\pi}
\cos\phi\sin\phi\dif\phi\dif\theta=\\=\frac 12\left(\left(1-\frac
d{\lambda_0}\right)e^{-d/\lambda_0}+\left(\frac d{\lambda_0}\right)^2
E_1(d/\lambda_0)\right).
\end{multline}
Thus, we find that the mean free path $\lambda$ at the distance $d$
away from the wall is given by the following scaling formula:
\begin{equation}
\label{eq:lambda_cosine}
\lambda_c(d)=\lambda_0\left(1-\frac 12 \beta_c(d/\lambda_0)\right),
\qquad\beta_c(x)=e^{-x}-x\beta_u(x).
\end{equation}
The formula above is the single-wall version of the scaling that was
tested by Guo et al. \cite{GuoShiZhe} (see Eq. (3) in \cite{GuoShiZhe}
with the distance to the second wall set to infinity), which, in turn,
was derived by Stops \cite{Stops} (see, for example, the first formula
on p.~689 in \cite{Stops}, and observe that it refers only to a half
of the total number of molecules).

\subsection{A comparison between the uniform and cosine scalings}

\begin{figure}
\includegraphics[width=\textwidth]{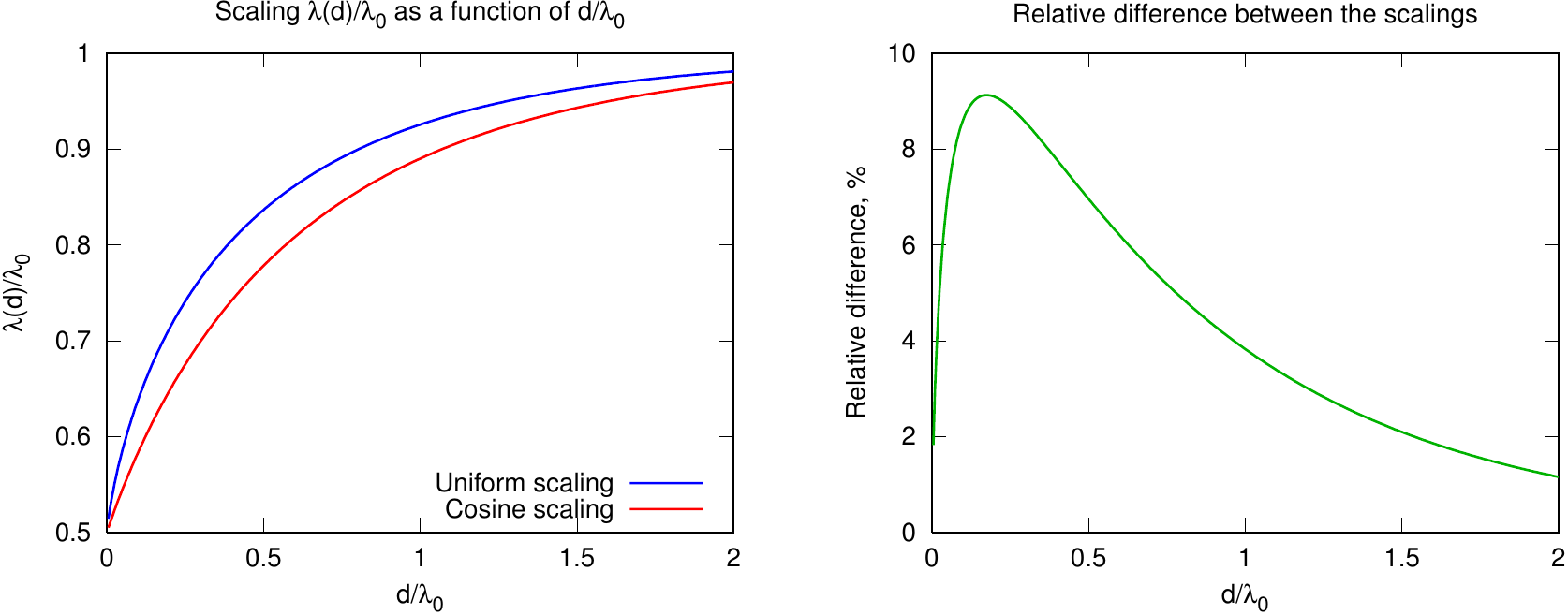}
\caption{A visual comparison of the uniform~\eqref{eq:lambda_uniform}
  and cosine~\eqref{eq:lambda_cosine} mean free path scalings.}
\label{fig:scaling}
\end{figure}

The difference between the uniform scaling
in~\eqref{eq:lambda_uniform} and the cosine scaling
in~\eqref{eq:lambda_cosine} is significant; we compare the two
scalings visually in Figure~\ref{fig:scaling}. Observe that even
though both scalings are monotonic concave functions which have the
same limits at zero and infinity, their behavior is otherwise
different.

The immediately noticeable qualitative difference between the two
scalings is the behavior of their derivatives at zero: while the
derivative of the uniform scaling in~\eqref{eq:lambda_uniform} at zero
is infinite, for the cosine scaling in~\eqref{eq:lambda_cosine} it
equals 1. This qualitative difference is important; as we show below,
the uniform scaling in~\eqref{eq:lambda_uniform} implies that, from
the fluid dynamics perspective, the flow is formally turbulent near
the wall boundary of the flow domain, whereas in the case of the
cosine scaling in~\eqref{eq:lambda_cosine} the second derivative of
the velocity remains finite at the wall boundary.

Quantitatively, the relative difference between the two scalings
reaches 9\% at the distance of roughly 20\% of the conventional mean
free path $\lambda_0$ away from the wall. It is also interesting that
the uniform scaling~\eqref{eq:lambda_uniform} in
Figure~\ref{fig:scaling} visually looks similar to the data computed
from the direct molecular dynamics simulations in~\cite{DonZhaRee}
(see Fig. 4 in~\cite{DonZhaRee}).

\section{An explicit Navier-Stokes solution for the shear flow near a
wall}

According to the kinetic theory of gases~\cite{ChaCow}, the viscosity
$\mu$ is proportional to the free mean path of a gas
molecule. Therefore, the uniform and cosine scalings for the mean free
path in~\eqref{eq:lambda_uniform} and~\eqref{eq:lambda_cosine},
respectively, imply that the viscosity scales in the same manner:
\begin{equation}
\label{eq:mu}
\mu(d)=\mu_0\left(1-\frac 12\beta(d/\lambda_0)\right),
\end{equation}
where $\mu_0$ is the conventional value of the viscosity (computed as
if no wall was present) for the given gas parameters, and $\beta$ is
either the uniform scaling $\beta_u$ from~\eqref{eq:lambda_uniform},
or the cosine scaling $\beta_c$ from~\eqref{eq:lambda_cosine}. The
same proposition was made in~\cite{GuoShiZhe}.

The transport equation for the velocity $\BV u$ is given by
\begin{equation}
\label{eq:velocity}
\parderiv{(\rho\BV u)}t+\div\left(\rho\BV u\BV u^T+p\BM I+\BM S
\right)=0,
\end{equation}
where $\rho$ is the density, $p$ is the pressure, and $\BM S$ is the
stress tensor. The transport equation for the stress tensor $\BM S$
with the linearized collision term is, in turn, given
by~\cite{Abr16,Gra,Mal}
\begin{equation}
\label{eq:stress}
\parderiv{\BM S}t+\div(\BV u\otimes\BM S+\BM Q)+\BM P\nabla\BV u+(\BM
P \nabla \BV u)^T+(1-\gamma)(\trace(\BM P\nabla\BV u)+\div\BV q)\BM
I=-\frac p\mu\BM S,
\end{equation}
where $\BM P=p\BM I+\BM S$ is the pressure tensor, $\BV q$ is the heat
flux, $\BM Q$ is the full centered third moment (a tensor of rank 3),
and $\gamma$ is the adiabatic exponent. Then, the Navier-Stokes
approximation for $\BM S$ is obtained under the assumption that all
terms in the left-hand side of~\eqref{eq:stress}, except for those
which involve $p$ and $\BV u$ only, are small and can be discarded.
This leads to the approximation
\begin{equation}
\label{eq:stress_NS}
\BM S=-\mu\left(\nabla\BV u+(\nabla\BV u)^T+(1-\gamma)(\div\BV u)\BM
I\right),
\end{equation}
which is then substituted into the velocity transport
equation~\eqref{eq:velocity} to obtain the usual Navier-Stokes
equation.

For a stationary shear flow, let us choose a coordinate system where
the $x$-coordinate points in the direction of the flow, and the
$y$-coordinate points in the direction of the shear. In this
situation, the velocity vector $\BV u$ points along the $x$-direction,
but the spatial dependence of the flow variables occurs in the
$y$-direction. As the $y$-coordinate measures the distance to the
wall, we will use $d$ to denote the $y$-coordinate:
\begin{equation}
\BV u=(u(d), 0, 0)^T,\quad \div\BM S=((\mu u')', 0, 0)^T,
\end{equation}
where the prime denotes the $d$-differentiation. This form of the
velocity yields, upon the substitution of the Navier-Stokes
approximation to $\div\BM S$ above in~\eqref{eq:stress_NS} back into
the stationary velocity equation~\eqref{eq:velocity},
\begin{equation}
\label{eq:shear_velocity}
(\mu u')'=0.
\end{equation}

\subsection{The velocity solution for the boundary value problem of
the shear flow}

For convenience, we set the flow velocity $u$ to zero at the wall
boundary of the flow domain, and to $u_0$ at the distance $d_0$ from
the boundary. Observe that in reality the gas flow velocity exhibits a
discontinuous ``slip'' directly at the wall surface
\cite{LilSad,LilSad2}, that is, the velocity of the gas flow at the
wall is not that of the wall itself. This kinetic effect, however, is
beyond the scope of the current work, as, from the perspective of the
fluid dynamics, we consider a spatial domain with suitable boundaries
(which do not have to be aligned with walls in general, as the domain
may simply be a bounded ``cutout'' of the free gas flow away from any
walls or other objects), and it is the properties of the gas flow
itself which are defined at the boundaries, irrespective of whether or
not a wall is present immediately outside the boundary.  Thus, in what
follows, we refer strictly to the velocity of the gas flow at a wall
boundary, regardless of the (possibly different) velocity of the wall
itself.

We further assume that the temperature of the gas flow varies
insignificantly, so that the conventional viscosity $\mu_0$
in~\eqref{eq:mu} (computed as if no wall was present) can be treated
as a constant.  Then, for a constant viscosity $\mu$ without the mean
free path scaling, the Navier-Stokes equation for the velocity of the
shear flow in~\eqref{eq:shear_velocity} reduces to
\begin{equation}
u''=0,
\end{equation}
which leads to the conventional linear shear velocity profile solution
\begin{equation}
\label{eq:linear_solution}
u(d)=u_0\frac d{d_0}.
\end{equation}
On the other hand, with the variable viscosity in~\eqref{eq:mu}, the
Navier-Stokes velocity equation~\eqref{eq:shear_velocity} for the
shear flow becomes
\begin{equation}
\label{eq:turbulent}
\left(\left(1-\frac 12\beta(d/\lambda_0)\right)u(d)'\right)'=0.
\end{equation}
Upon the integration with the same boundary conditions as above,
\eqref{eq:turbulent} results in
\begin{equation}
\label{eq:nonlinear_solution}
u(d)=u_0\frac{b(d/\lambda_0)}{b(d_0/\lambda_0)},\qquad b(x)=\int_0^x
\frac{\dif y}{2-\beta(y)},
\end{equation}
where $\beta$ is given by either the uniform scaling
in~\eqref{eq:lambda_uniform}, or the cosine scaling
in~\eqref{eq:lambda_cosine}.

\subsection{Wall boundary turbulence in the case of the uniform
viscosity scaling}

Observe that, in the case of the uniform viscosity scaling (that is,
$\beta=\beta_u$ from~\eqref{eq:lambda_uniform}) the identity above
in~\eqref{eq:turbulent} requires the second derivative $u''(d)$ to
become infinite at the wall boundary. Indeed,
substituting~\eqref{eq:lambda_uniform} into~\eqref{eq:turbulent}, we
obtain
\begin{equation}
u(d)''=\frac{u(d)'}{2\lambda_0}\left(1-\frac 12\beta_u(d/\lambda_0)
\right)^{-1}\beta_u'(d/\lambda_0).
\end{equation}
However, observe that
\begin{equation}
\beta_u'(x)=\left(e^{-x}-x E_1(x)\right)'=-E_1(x),
\end{equation}
and that $E_1(0)$ is infinite. Thus, $u''(0)$ must also become
infinite for~\eqref{eq:turbulent} to hold. Observe that this condition
does not violate the assumptions made above on the stress
equation~\eqref{eq:stress}, as the latter does not involve the second
derivatives of the velocity, and the Navier-Stokes approximation of
the stress is still valid. In contrast, the cosine mean free path
scaling in~\eqref{eq:lambda_cosine} does not have this property.

Recall that sufficiently large second derivatives of the velocity $u$
cause the viscous term in the Navier-Stokes velocity equation
in~\eqref{eq:velocity} to become of a comparable order in magnitude to
the transport terms (as under normal conditions the viscous term is
usually small due to low viscosity), which is what is observed in a
turbulent flow. In fact, based on this observation,
Boussinesq~\cite{Bou} proposed to model the effect of the large second
derivatives of the velocity in a turbulent flow with the ``eddy
viscosity'', that is, an artificially large viscosity coefficient to
compensate for the unresolved turbulent fine-grained motion of the
flow and to achieve a similar net effect of the stress divergence. The
same principle is used in various numerical models of turbulent
flow~\cite{Wilc}. Thus, from the fluid dynamics perspective, the
infinite second derivative of the velocity at the wall boundary
formally makes the boundary layer of the flow turbulent.

Observe, however, that in the current setting the turbulence in the
boundary layer is the result of the imposed condition that the viscous
term in the velocity equation~\eqref{eq:velocity} is bounded near the
wall, which is directly opposite to the effect of the turbulence away
from the wall.  Nonetheless, this effect suggests that the uniform
mean free path scaling in~\eqref{eq:lambda_uniform} may hint at the
actual physical mechanism of the turbulence manifestation in a shear
flow -- for example, when some appropriate instability conditions are
satisfied, the turbulence near the wall boundary may ``spill'' into
the flow outside the boundary layer.

\section{Numerical experiments}

Here we compare the gas flow velocity profiles corresponding to the
uniform mean free path scaling in~\eqref{eq:lambda_uniform} and the
cosine scaling in~\eqref{eq:lambda_cosine} with the Direct Simulation
Monte Carlo computations (DSMC) for argon and nitrogen. For a better
reliability of the DSMC computations results, we used two different
publicly available DSMC software codes: one is the
DS1V~\cite{Bird}\footnote{Available at
  \href{http://www.gab.com.au}{http://www.gab.com.au}}, and another is
the dsmcFoam~\cite{ScaRooWhiDarRee}\footnote{Part of the OpenFOAM
  software,
  \href{http://www.openfoam.org}{http://www.openfoam.org}}. For both
argon and nitrogen, the temperature and pressure were set at about
15$^\circ$ C and 100 kPa respectively, which correspond to the normal
conditions at sea level. The shear in the flow was imposed by placing
two infinite parallel moving walls (the so-called Couette flow) with
diffuse reflection boundary conditions at the distance of 1000
nanometers from each other. The relative difference in the wall
velocity was 100 meters per second.  Due to the symmetry of the flow,
below we consider only a half of the flow domain, between one of the
walls and the middle point between the walls (such that the width of
our domain is 500 nm).

For convenience, we removed the wall slip from the DSMC velocity
profile by subtracting its zero-distance value. This also resulted in
the DSMC velocity at 500 nm away from the wall being about 45 m/s
(rather than half of the wall velocity difference, that is, 50 m/s).
As explained above, we did not consider the wall itself to be a part
of the domain, and thus chose the coordinate system so that the DSMC
flow velocity was zero at the wall boundary of the domain.

The observed temperature variation in the DSMC solutions was about two
degrees, which corresponds to the variation of about 0.3\% in the
conventional value of the viscosity $\mu_0$ across the flow domain.
The effect of the velocity shear in the DSMC simulation on the assumed
uniformity of free flight directions is also small: indeed, observe
that the change of 45 m/s on the 500 nm scale translates to about 6
m/s on the 60-70 nm (mean free path) scale. At the same time, average
speeds of molecules at the room temperature range from about 400
(argon) to about 500 (nitrogen) m/s, so that the velocity shear effect
accounts for no more than 1.5\% of the typical molecular velocity on
the mean free path scale.

\begin{figure}
\includegraphics[width=\textwidth]{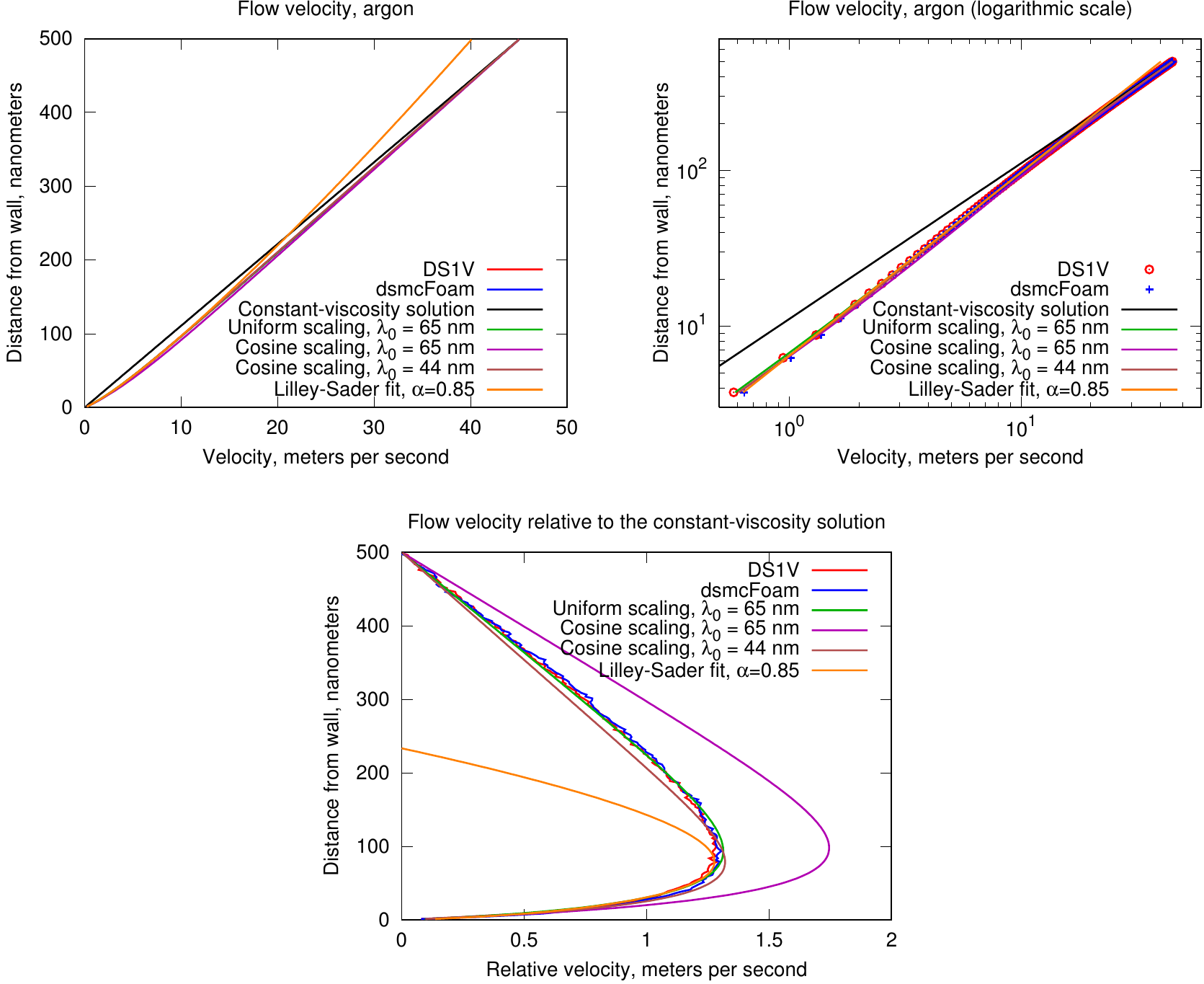}
\caption{The velocity of the shear flow for argon.}
\label{fig:argon}
\end{figure}

\begin{figure}
\includegraphics[width=\textwidth]{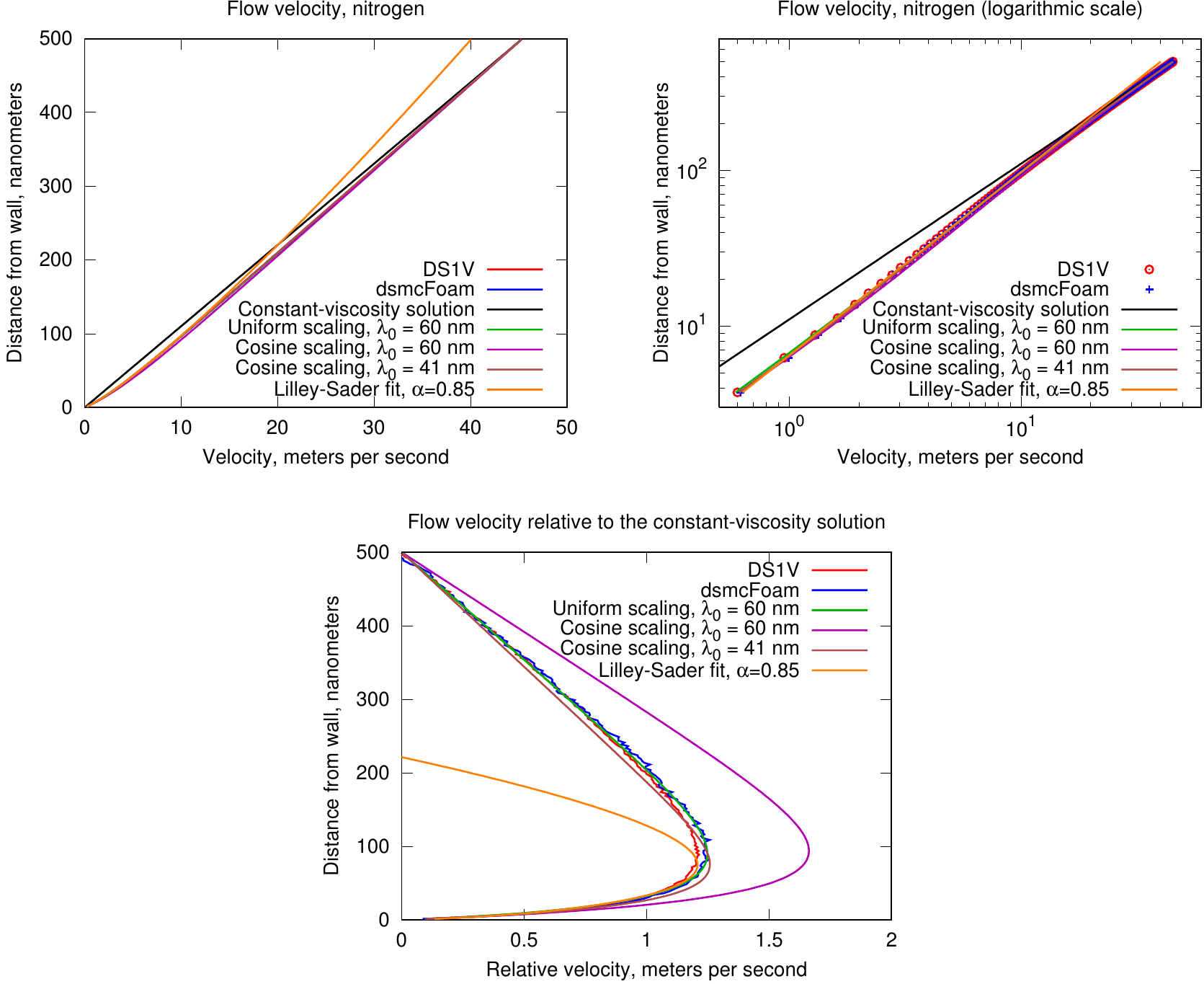}
\caption{The velocity of the shear flow for nitrogen.}
\label{fig:nitrogen}
\end{figure}

In Figures~\ref{fig:argon} and~\ref{fig:nitrogen} we compare the
constant-viscosity linear solution in~\eqref{eq:linear_solution} and
the nonlinear solutions in~\eqref{eq:nonlinear_solution} corresponding
to the uniform mean free path scaling in~\eqref{eq:lambda_uniform} and
cosine scaling~\eqref{eq:lambda_cosine} for argon and nitrogen with
the corresponding velocity profiles computed by the DS1V~\cite{Bird}
and dsmcFoam~\cite{ScaRooWhiDarRee} software as described above. We
used the DSMC-computed velocity values at 500 nm as the boundary
conditions $u_0$ in~\eqref{eq:linear_solution} and
\eqref{eq:nonlinear_solution}.  We set the conventional mean free path
$\lambda_0$, which is a parameter in~\eqref{eq:nonlinear_solution}, to
$65$ nanometers for argon (Figure~\ref{fig:argon}) and to $60$
nanometers for nitrogen (Figure~\ref{fig:nitrogen}), which appear to
be realistic enough for the normal
conditions~\cite{HirCurBir}. Observe that the range of the flow domain
(from zero to 500 nm) extends well beyond the Knudsen boundary layer
region.

For illustrative purposes, we also show the velocity power fit of
Lilley and Sader \cite{LilSad,LilSad2} of the form
\begin{equation}
\label{eq:Lilley_Sader}
u(d)=C(d/\lambda_0)^{\alpha},
\end{equation}
where the constant $\alpha$ was computed by the least squares fit on
the DSMC velocity data between the wall and $\lambda_0$ (for both
argon and nitrogen, the least squares algorithm yielded
$\alpha=0.85$). The constant coefficient $C$ was set to the average
value of DS1V and dsmcFoam flow velocities at the distance of the
conventional mean free path $\lambda_0$ from the wall.

Observe that the Knudsen boundary layers, produced by the DSMC
simulations, are captured quite accurately by the velocity solution
in~\eqref{eq:nonlinear_solution} with the uniform mean free path
scaling in~\eqref{eq:lambda_uniform} for both argon and nitrogen. In
contrast, the conventional constant-viscosity Navier-Stokes
solution~\eqref{eq:linear_solution} fails to develop a boundary layer
at all, while the solution for the cosine scaling
in~\eqref{eq:lambda_cosine} develops a noticeably stronger Knudsen
layer than those produced by the DSMC methods. The Lilley-Sader
velocity power fit~\eqref{eq:Lilley_Sader} is accurate within the
Knudsen boundary layer range (which is where it was fitted), but
quickly diverges from the DSMC solutions outside the Knudsen boundary
layer.

Additionally, we computed the velocity profiles using the cosine
scaling in~\eqref{eq:lambda_cosine} with the mean free path parameter
$\lambda_0$ artificially chosen so as to match the resulting velocity
profile to the DSMC results as close as possible (44 nm for argon, and
41 nm for nitrogen). The results are also shown in
Figures~\ref{fig:argon} and~\ref{fig:nitrogen}. Observe that, even
given the unrealistic values of $\lambda_0$ for an artificially better
fit, the resulting velocity profiles are still not as accurate as the
ones produced by the uniform scaling in~\eqref{eq:lambda_uniform} with
realistic values of $\lambda_0$.

We summarize the computational results in Table~\ref{tab:errors},
which contains relative errors (in the sense of the usual Euclidean
square norm) between the averaged DSMC velocity profile and different
scaling approximations in the full flow domain between the wall
boundary and 500 nm distance away from the wall. Observe that the
uniform scaling in~\eqref{eq:lambda_uniform} offers better accuracy
than the cosine scaling in~\eqref{eq:lambda_cosine}.
\begin{table}
\begin{tabular}{|c|c|}
\hline
\begin{tabular}{c}
Argon \\
\hline
\begin{tabular}{c|c}
Scaling & Relative error \\
\hline\hline
Uniform, $\lambda_0=65$ nm & $7.564 \cdot 10^{-4}$ \\
Cosine, $\lambda_0=65$ nm & $1.1 \cdot 10^{-2}$ \\
Cosine, $\lambda_0=44$ nm & $1.822 \cdot 10^{-3}$ \\
\end{tabular}
\end{tabular}
&
\begin{tabular}{c}
Nitrogen \\
\hline
\begin{tabular}{c|c}
Scaling & Relative error \\
\hline\hline
Uniform, $\lambda_0=60$ nm & $5.399 \cdot 10^{-4}$ \\
Cosine, $\lambda_0=60$ nm & $1.059 \cdot 10^{-2}$ \\
Cosine, $\lambda_0=41$ nm & $1.444 \cdot 10^{-3}$ \\
\end{tabular}
\end{tabular}
\\ \hline
\end{tabular}

\vspace{1EX}

\caption{Relative errors for different scalings.}
\label{tab:errors}
\end{table}

\section{Concluding remarks}

In the current work we propose a new mean free path (and,
subsequently, viscosity) scaling near a wall based on the uniform
distribution of the molecular free flight directions incident on the
wall. We also derive the corresponding velocity solution to the
conventional Navier-Stokes equation for the shear flow near a wall,
based on the proposed scaling. We compare the proposed uniform scaling
with the scaling based on the cosine distribution of the incident
flight directions suggested by Stops~\cite{Stops} and tested by Guo et
al.~\cite{GuoShiZhe}, as well as the Lilley-Sader velocity power
fit~\cite{LilSad,LilSad2} for the shear flow of argon and nitrogen at
normal conditions. The Direct Simulation Monte Carlo computations of
two different software codes, DS1V~\cite{Bird} and
dsmcFoam~\cite{ScaRooWhiDarRee} are used for the validation of the
results. The proposed uniform scaling is found to be more accurate
than the other studied approximations in the domain of the simulated
flow, which extends well beyond the Knudsen boundary layer.

The proposed viscosity scaling also appears to be convenient for the
computational fluid dynamics modeling, since it can be readily
integrated into the existing models of a viscous gas flow, provided
that the value of the conventional mean free path $\lambda_0$ can be
reliably estimated from the macroscopic parameters of the gas
dynamics. Suitable approximate formulas for the exponential integral
(which is a part of the mean free path and viscosity scaling formula),
can be found in~\cite{Gia}.

An unexpected finding of this work is the inherent presence of
turbulence, in the formal fluid dynamics sense, within the Knudsen
boundary layer near the wall for the uniform viscosity scaling. It is
interesting that this result is purely analytic; clearly, the DSMC
simulations or experimental measurements cannot detect it. This result
could potentially provide some hints at the physical mechanism behind
the visible manifestation of turbulence outside the boundary layer --
indeed, if the turbulence covertly exists within the Knudsen boundary
layer, then it could be a matter of suitable conditions for it to
spread from the boundary layer into the outside flow. Obviously, this
finding necessitates a further study.

\ack

The work was supported by the Office of Naval Research grant
N00014-15-1-2036.

\end{document}